\documentclass[amsmath,amssymb,aps,prl,reprint,superscriptaddress,footnoteinbib]{revtex4-1}
\usepackage{amsmath,amssymb}
\usepackage[english]{babel}
\usepackage{ bbold }
\usepackage{upgreek}
\usepackage{dsfont}
\usepackage{graphicx}
\usepackage{comment}
\usepackage{braket}
\usepackage{colortbl}
\usepackage{blindtext}
\usepackage{hyperref}
\begin{document}

\title{Cooling Nanorotors by Elliptic Coherent Scattering}
	
\author{Jonas Sch\"afer}
\affiliation{Faculty of Physics, University of Duisburg-Essen, Lotharstra\ss e 1, 47048 Duisburg, Germany}

\author{Henning Rudolph}
\affiliation{Faculty of Physics, University of Duisburg-Essen, Lotharstra\ss e 1, 47048 Duisburg, Germany}

\author{Klaus Hornberger}
\affiliation{Faculty of Physics, University of Duisburg-Essen, Lotharstra\ss e 1, 47048 Duisburg, Germany}

\author{Benjamin A. Stickler}
\affiliation{Faculty of Physics, University of Duisburg-Essen, Lotharstra\ss e 1, 47048 Duisburg, Germany}
\affiliation{QOLS, Blackett Laboratory, Imperial College London, London SW7 2AZ, United Kingdom}

\begin{abstract}
Simultaneously cooling the rotational and translational motion of nanoscale dielectrics into the quantum regime is an open task of great importance for sensing applications and quantum superposition tests. Here, we show that the six-dimensional ground state can be reached by coherent-scattering cooling with an elliptically polarized and shaped optical tweezer. We determine the  cooling rates and steady-state occupations in a realistic setup, and discuss applications for mechanical sensing and fundamental experiments.
\end{abstract}

\maketitle

{\it Introduction---} Optically levitating nanoparticles in ultra-high vacuum yields an unprecedented degree of environmental isolation \cite{millen2020optomechanics}, rendering these systems ideally suited for precision sensing \cite{chaste2012,ranjit2016,hempston2017,ahn2020} and for mesoscopic quantum superposition tests \cite{romero2011,bateman2014,arndt2014,wan2016,kaltenbaek,stickler2018,rudolph2020,rakhubovsky2020}. For {several} applications it is crucial to cool the rotational and translational particle dynamics into the quantum regime. Recently, center-of-mass ground-state cooling has been achieved \cite{delic2020} using the method of coherent-scattering cooling \cite{delic2019part2,windey2019}. In this work, we show that elliptically polarized and shaped tweezers enable cooling of nanoparticles into their simultaneous rotational and translational ground state.

The setup of coherent-scattering cooling consists of an optical tweezer levitating a nanoparticle inside a high-finesse cavity. If the tweezer is slightly red detuned from the cavity resonance, the particle  motion loses energy by scattering tweezer photons into the cavity mode \cite{salzburger2009,gonzalez2019}. In contrast to conventional optomechanical setups \cite{chang2010,romero2010,barker2010,asenbaum2013,kiesel2013,millen2015,stickler2016,fonseca2016}, the cavity mode is nearly empty in the system's steady state, leading to a significant reduction of laser phase noise \cite{delic2019part2} and holding the prospect of reaching the strong coupling regime \cite{sommer2020}.

What distinguishes levitated nanoparticles from other optomechanical systems is their ability to rotate. Desired or not, rotations of any levitated object must be controlled to fully exhaust its potential for sensing \cite{moore2014,rider2016} and for future quantum superposition tests \cite{scala2013,bateman2014,wan2016,kaltenbaek,knobloch2017}. The classical rotation dynamics of nanoscale objects can be manipulated with linearly and circularly polarized lasers \cite{hoang2016,kuhn2017,ahn2018,reimann2018,vanderlaan2020}, enabling high-precision pressure \cite{kuhn2017part2} and torque \cite{ahn2020} sensing. In the quantum regime, the non-harmonicity  of the rotational spectrum gives rise to pronounced interference effects  \cite{stickler2018,ma2020} and might be useful for quantum information processing \cite{grimsmo2020,albert2019}.

Recent experiments demonstrate rotational cooling \cite{delord2020, bang2020} by aligning particles in a space-fixed direction and damping their librations around this axis. While such schemes allow reaching the quantum regime with ultra-thin rod-shaped objects \cite{stickler2016,seberson2019}, they cease to be efficient once the particle shape significantly deviates from this idealization and rotational precession \cite{rashid2018} becomes important. Specifically, rotations around the axis of maximal susceptibility experience only weak cooling but strongly influence the libration dynamics \cite{bang2020}. Finding a strategy to cool an arbitrarily shaped particle through the non-harmonic regime of rotations is therefore a prerequisite for future quantum applications of levitated objects.

\begin{figure}[b]
	\centering
	\includegraphics[width=1\linewidth]{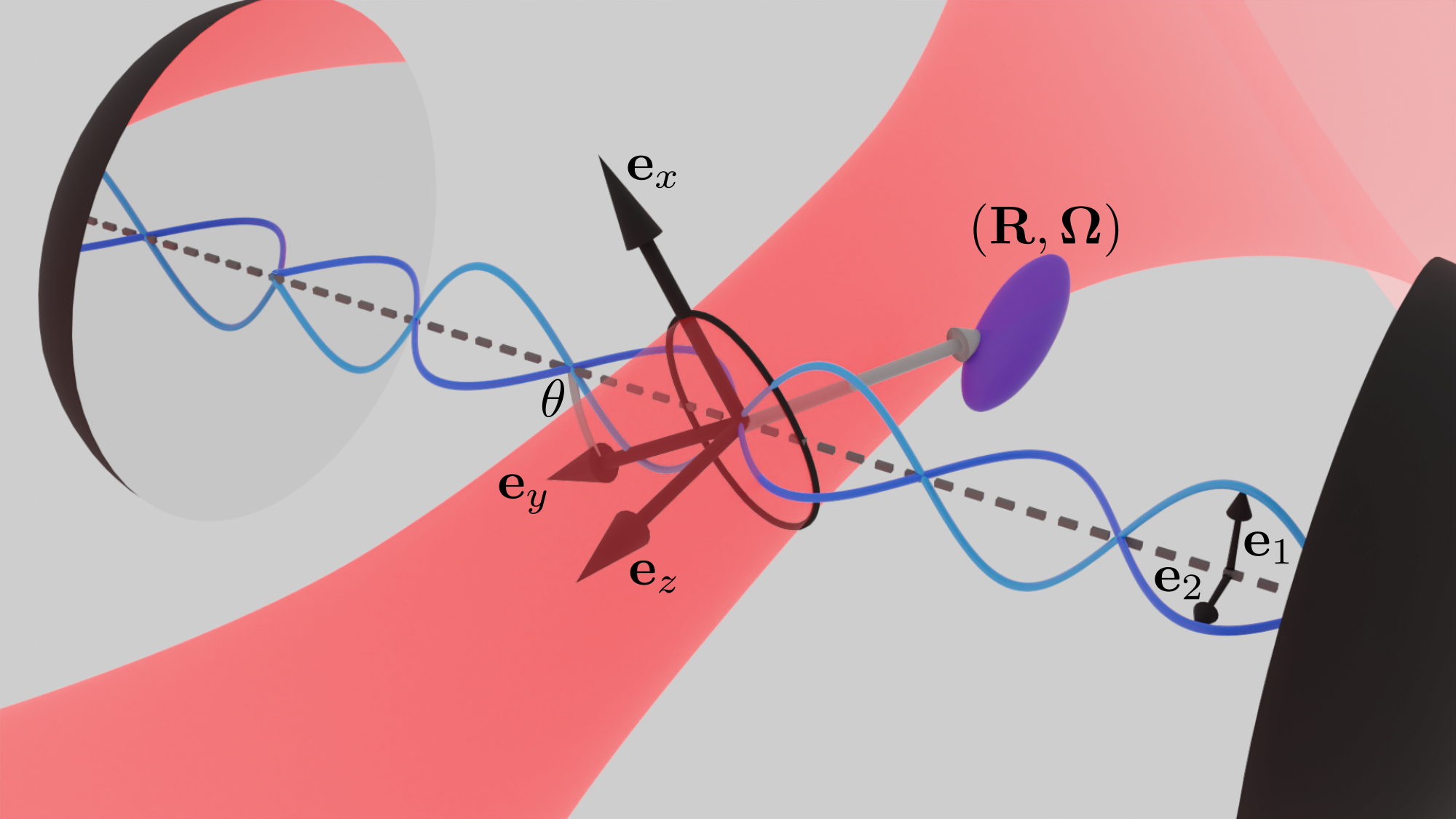}
	\caption{An aspherical nanoparticle (purple ellipsoid) is trapped by a tweezer (red) inside an optical cavity with the principal axis (dashed line) orthogonal to the tweezer propagation direction. The tweezer is elliptically polarized and exhibits an elliptic intensity profile. The induced  polarization field depends on the  particle center-of-mass position ${\bf R}$ and orientation $\Omega$, driving two cavity modes (blue lines) with orthogonal polarizations $\mathbf{e}_{1,2}$. The resulting coupled nanoparticle-cavity dynamics can cool the nanoparticle motion into the 6D ro-translational quantum ground state.}
	\label{fig:sketch54}
\end{figure}

In the present Letter and the accompanying article \cite{prasubmission}, we demonstrate how the non-linearity of rotation dynamics can be overcome by coherent-scattering cooling with an elliptically polarized and shaped tweezer, see Fig. \ref{fig:sketch54}. The elliptical polarization of the light field introduces two space-fixed axes to control the rotations \cite{stapelfeldt2003}, rather than a single polarization axis. When placed inside a cavity, the nanoparticle rotations couple to two orthogonally polarized cavity modes, which in turn cool different mechanical degrees of freedom. For suitably chosen optical parameters, the resulting coupling cools sub-wavelength aspherical objects into their joint 6D ro-translational quantum ground state if they are close to spherical.  We calculate the cooling rates and steady-state occupations for experimentally realistic situations, show that the expected torque sensitivities surpass state-of-the-art estimates by several orders of magnitude, and discuss how this setup can be used to generate ultra-fast spinning, ultra-cold nanoparticles.

{\it Nanoparticle-light interaction---} We consider an ellipsoidal particle of mass $m$, with three different principal-axis diameters $\ell_a<\ell_b<\ell_c$ and associated moments of inertia $I_a>I_b>I_c$. For sub-wavelength particles, the internal field is described by a linear susceptibility tensor $\chi(\Omega) = R(\Omega)\chi_0R^{ T}(\Omega)$ (Rayleigh-Gans approximation), where  $R(\Omega)$ rotates from the space-fixed frame to the  principal-axes frame and $\Omega = (\alpha,\beta,\gamma)$ denotes the Euler angles in the $z$-$y'$-$z''$-convention. The susceptibility tensor $\chi_0 = \text{diag}(\chi_a,\chi_b,\chi_c)$ contains the susceptibilities along the ellipsoid's principal axes. They can be calculated in terms of elliptic integrals \cite{hulst1981,prasubmission}, yielding $\chi_a < \chi_b < \chi_c$.

For a given laser field $\mathbf{E}(\mathbf{r})e^{-i \omega t}$, the time-averaged force and torque acting on the particle are  dominated by the conservative optical potential
\begin{equation}\label{eq:Vopt}
V_{\rm opt} (\mathbf{R},\Omega) = - \frac{\varepsilon_0 V}{4} \mathbf{E}^* (\mathbf{R})\cdot \chi(\Omega)\mathbf{E} (\mathbf{R}),
\end{equation}
to first order of the particle volume $V$ \cite{stickler2016}, with the center-of-mass position $\mathbf{R}=(x,y,z)$. Thus, if the electric field is linearly polarized, the optical potential tends to align the particle axis of maximal susceptibility with the field polarization.

In addition to the conservative potential \eqref{eq:Vopt}, the laser also exerts a radiation pressure force and torque. The latter follow from the electric-field integral equation as \cite{prasubmission}
\begin{align} \label{eq:force}
 {\bf F}_{\rm rad} =   \frac{\varepsilon_0 k^3 V^2}{12\pi}\text{Im} \left[ (\chi\mathbf{E}^*) \cdot [\nabla \otimes (\chi\mathbf{E})]^T \right],
\end{align}
and
\begin{align} \label{eq:torque}
\mathbf{N}_{\rm rad} = \frac{\varepsilon_0 k^3 V^2}{12\pi} {\rm Im}\left[ \left( \chi^2 \mathbf{E}^* \right)\times \mathbf{E} -  \left( \chi \mathbf{E}^*\right) \times \left( \chi \mathbf{E}\right)\right],
\end{align}
where we omitted the dependence on ${\bf R}$ and $\Omega$. Force and torque are proportional to the  cubed wavenumber $k$ and to the squared particle volume, and are consistent with Ref.~\cite{stickler2016} for symmetric objects. An optical torque consistent with \eqref{eq:torque} has been experimentally observed to  accelerate nanorotors up to GHz frequencies \cite{kuhn2017,kuhn2017part2,reimann2018,ahn2018}; it vanishes for linear field polarization.

{\it Coupled nanoparticle cavity-dynamics---} If the particle is trapped inside a high-finesse optical cavity, the mechanical motion can couple  strongly to two near-degenerate cavity modes, see Fig. \ref{fig:sketch54}.  This interaction is determined by the total field, 
\begin{align}
\mathbf{E}(\mathbf{r}) = \sqrt{\frac{2\hbar\omega}{\varepsilon_0 V_{\rm c}}} \left [\epsilon \mathbf{e}_{\rm t} f_{\rm t}(\mathbf{r})  + \sum_{j=1,2} b_j \mathbf{e}_j f_{\rm c}(\mathbf{r}) \right ],
\end{align}
where $\omega$ denotes the tweezer frequency, $V_{\rm c}$ is the cavity mode volume, $\epsilon$ and $b_{1,2}$ are the dimensionless tweezer and cavity amplitudes, ${\bf e}_{\rm t}$ and ${\bf e}_{1,2}$  the corresponding polarization vectors, and $f_{\rm t}({\bf r})$ and $f_{\rm c}({\bf r})$  the mode functions.

The cavity resonance frequency $\omega_{\rm c}$ is detuned from the tweezer frequency by $\Delta = \omega - \omega_{\rm c}$. Near the cavity axis, the mode function can be approximated as a standing wave $f_{\rm c}(\mathbf{r})= \cos\left[k\left( \mathbf{e}_2 \times \mathbf{e}_1 \right)\cdot \mathbf{r} + \phi \right]$, where  $\phi$ describes the relative positioning of cavity and tweezer. The Gaussian envelope of the cavity modes with cavity waist $w_{\rm c}$ and length $L$ determines the mode volume $V_{\rm c}$, but can be neglected for the dynamics.

The tweezer quadrature $\epsilon$ is determined by the tweezer power and will be chosen real and positive. The tweezer mode function can be approximated by a traversing Gaussian beam with an elliptic intensity profile, $f_{\rm t}(\mathbf{r}) = \exp[-(x^2/w_x^2 + y^2/w_y^2)/r^2(z)] e^{i[kz - \phi_{\rm t}(\mathbf{r})]}/r(z)$ with the beam waists $w_{x,y}$. The latter determine the Rayleigh range $z_{\rm R}$, which in turn sets the broadening factor $r(z)$ and the the Gouy phase $\phi_{\rm t}({\bf r})$ \cite{gonzalez2019,prasubmission}. 

We assume the tweezer propagation direction $\mathbf{e}_z$ to be  orthogonal to the cavity axis, with  $\theta$ the angle between the latter and the $y$-axis, see Fig.~\ref{fig:sketch54}. The cavity mode polarizations can be chosen as $\mathbf{e}_1 = \cos\theta {\bf e}_x -\sin\theta {\bf e}_y$ and $\mathbf{e}_2 = \mathbf{e}_z$. In order to achieve trapping in all three orientational degrees of freedom, we choose the tweezer to be elliptically polarized,  $\mathbf{e}_{\rm t} = \cos\psi\, \mathbf{e}_{{\rm t},1} + i\sin\psi\, \mathbf{e}_{{\rm t},2}$ with the ellipticity $\psi \in [0,\pi/4]$. The two polarization axes are rotated with respect to the tweezer main axes by the angle $\zeta$, ${\bf e}_{{\rm t},1} = \cos \zeta {\bf e}_x - \sin \zeta {\bf e}_y$ and ${\bf e}_{{\rm t},2} = \sin \zeta {\bf e}_x + \cos \zeta {\bf e}_y$.

The nanoparticle equations of motion are given by the optical potential \eqref{eq:Vopt} together with the non-conservative force \eqref{eq:force} and torque \eqref{eq:torque}.  Taking the finite cavity line width $\kappa$ into account, the dynamics of the cavity modes $b=(b_1,b_2)$ are described by $\dot b = \left ( i\Delta_{\rm eff} - \kappa_{\rm eff} \right )b + \eta$, with the vector $[\eta ]_j = -\epsilon  \mathbf{e}_j\cdot (i U_0 \chi  + \gamma_{\rm sc} \chi^2 /2)\mathbf{e}_{\rm t} f_{\rm c} f_{\rm t} $, the coupling frequency $U_0 = -\omega V/2 V_{\rm c}$ and the Rayleigh scattering rate $\gamma_{\rm sc} = \omega k^3 V^2/6\pi V_{\rm c}$. The matrices $\Delta_{\rm eff}$ and $\kappa_{\rm eff}$ describe the effective detuning and linewidth for a given particle position and orientation,
\begin{align} \label{eq:delta}
\left[\Delta_{\rm eff}\right]_{jj'} = \Delta \delta_{jj'} - U_0 f_{\rm c}^2 \mathbf{e}_j \cdot \chi \mathbf{e}_{j'},
\end{align}
and
\begin{align}
	[\kappa_{\rm eff}]_{jj'} = \kappa\delta_{jj'} + \frac{\gamma_{\rm sc}}{2}f_{\rm c}^2 \mathbf{e}_j \cdot \chi^2 \mathbf{e}_{j'}.
\end{align}
It can be demonstrated that the combined rotational and translational particle motion is always cooled down if the tweezer is sufficiently far red-detuned, i.e. for $ \Delta < U_0 \chi_c$  \cite{prasubmission}. In this limit, the potential is always stiffer when the particle moves up the potential slope than when it moves toward the minimum.

{\it Deep trapping regime---} We harmonically expand the nanoparticle-cavity dynamics around the tweezer potential minimum $q_{\rm tw} = ({\bf R}_{\rm tw}, \Omega_{\rm tw})$, where the axis with the largest (intermediate) susceptibility aligns in parallel to the stronger (weaker) tweezer polarization axis ${\rm Re}(\mathbf{e}_{\rm t})$ (${\rm Im}(\mathbf{e}_{\rm t})$) \cite{stapelfeldt2003}. At the minimum $\mathbf{R}_{\rm tw}= (0,0,0)$ and $\Omega_{\rm tw} = (-\zeta,\pi/2,0)$ the cavity modes attain the amplitudes $b_{\rm tw}$, implying that the effective detuning matrix \eqref{eq:delta} is diagonal and that mode $b_2$ is empty \cite{prasubmission}.

The coupling to the cavity modes as well as the non-conservative tweezer torque displace the equilibrium configuration $(q_{\rm eq}, b_{\rm eq})$ slightly away from the tweezer minimum $(q_{\rm tw},b_{\rm tw})$. For small offsets, this does not affect the dynamics of the small deviations $\delta b = b - b_{\rm eq}$ and of the small mechanical displacements described by the mode operators $a_q = (q - q_{\rm eq} + i p_q/m_q \omega_q) /2 q_{\rm zp}$. Here, the zero-point amplitude $q_{\rm zp} = \sqrt{\hbar/2m_q \omega_q}$ is determined by the {Hamiltonian $H_0$ of the free translational and rotational motion \cite{prasubmission}, in terms of the} effective masses $m_q^{-1} = \partial_{p_q}^2 H_0(q_{\rm tw})$ and by the trapping frequencies $\omega_q^2 = \partial^2_q V_{\rm opt}(q_{\rm tw},b_{\rm tw})/m_q$. The opto-mechanical coupling frequencies follow from the interaction potential \eqref{eq:Vopt} as $g_{jq} = - q_{\rm zp}\partial_{b_j} \partial_q V_{\rm opt}(q_{\rm tw},b_{\rm tw})/\hbar$. These couplings imply that the optical mode $\delta b_1$ only interacts with coordinates that shift the effective detuning \eqref{eq:delta} along the ${\bf e}_1$-axis, i.e. $q \in S_1 = \{x,y,z,\alpha\}$. In a similar fashion, mode $\delta b_2$ only couples to $q \in S_2 = \{\beta, \gamma\}$. The equilibrium orientations of $S_2$ coincide with their minimal values in the tweezer potential since $b_2$ vanishes in the steady state. 

The resulting total Hamiltonian decomposes into two non-interacting contributions, $H = H_1 + H_2$, with
\begin{align} \label{eq:ham2}
\frac{H_j}{\hbar} = &\sum_{q \in S_j} \omega_q a_q^\dagger a_q -\sum_{q q' \in S_j} g_{qq'} (a_q + a_q^\dagger)(a_{q'} + a_{q'}^\dagger) \nonumber\\
& -\sum_{ q \in S_j}\left[g_{jq} \delta b_j \left(a_q + a_q^\dagger\right) + \text{h.c.}\right]-\Delta_j \delta b_j^\dagger \delta b_j.
\end{align}
Here, the detunings $\Delta_j = [\Delta_{\rm eff}(q_{\rm tw})]_{jj}$ follow from \eqref{eq:delta} and the cavity-mediated mechanical coupling rates are $g_{qq'} = -q_{\rm zp} q'_{\rm zp} \partial_q \partial_{q'} V_{\rm opt}(q_{\rm tw}, b_{\rm tw})/\hbar$.

\begin{table}[b]
	\centering
	\caption{Stationary phonon occupations for three silicon ellipsoids with principal axes $(\ell_a,\ell_b,\ell_c)$ in the weak-coupling approximation. Room temperature (r.t.) indicates that the respective degree of freedom is not cooled in the deep trapping regime. The occupations are obtained for a $1550\,$nm cavity, at ellipticity angle $\psi=\pi/6$, $\zeta=0$ and $10^{-9}\,$mbar. All other parameters were chosen to achieve efficient cooling. (70,70,70): cavity length $L=3\,$mm, cavity waist $w_c=40\,\mu$m, linewidth $\kappa=300\,$kHz, tweezer power $P=0.5\,$W, waists $w_x=1.6\,\mu$m, $w_y=1.3\,\mu$m,  detuning $\Delta=-500\,$kHz,  $\theta=\pi/4$, and $\phi=3\pi/8$. (25,40,100) first row: same as for (70,70,70), except $P=0.1$\,W, $\kappa=2\,$MHz, $\Delta=-11\,$MHz, $\theta=\pi/2$ and $\phi=0$. (25,40,100) second row: same as for (70,70,70). (69,70,71): $L=1.5\,$mm, $w_c=30\,\mu$m, $\kappa=600\,$kHz, $P=0.1\,$W, $w_x=800\,$nm, $w_y=650\,$nm, $\Delta=-1.8\,$MHz, $\theta=\pi/4$, and $\phi=3\pi/8$. The steady-state occupations for the $x'-$ and $y'-$modes are approximate because they are not in the strict weak-coupling regime for the given parameters.}
	\label{tab:occ}
	\begin{tabular}{c|cccc|cc}
	\hline \hline
	 particle shape & \multicolumn{4}{c|}{cooled by $b_1$} & \multicolumn{2}{c}{cooled by $b_2$} \\ 
		 $(\ell_a,\ell_b,\ell_c)$ [nm]& $n_{x'}$ & $n_{y'}$ & $n_{z'}$ & $n_{\alpha'}$ & $n_{\beta'}$ & $n_{\gamma'}$ \\[0.3cm]
		\hline
		$(70,70,70)$ & 0.1 & 0.1 & 0.5 & r.t. & r.t. & r.t. \\[0.3cm]
		\hline
		$(25,40,100)$ & r.t. & r.t. & $\gtrsim 10^2$ & 0.1 & 0.2 & 0.1 \\
		& 0.2 & 0.3 & 0.4 & $\gtrsim 10^3$ & $\gtrsim 10^3$ & $\gtrsim 10^3$ \\[0.3cm]
		\hline
		$(69,70,71)$ & $0.1$ & $0.1$ & 0.9 & 0.3 & 0.9 & 0.2 \\[0.3cm] 
		\hline \hline
	\end{tabular}
\end{table}

\begin{figure*}[t]
    \centering
    \includegraphics[width = 0.99\textwidth]{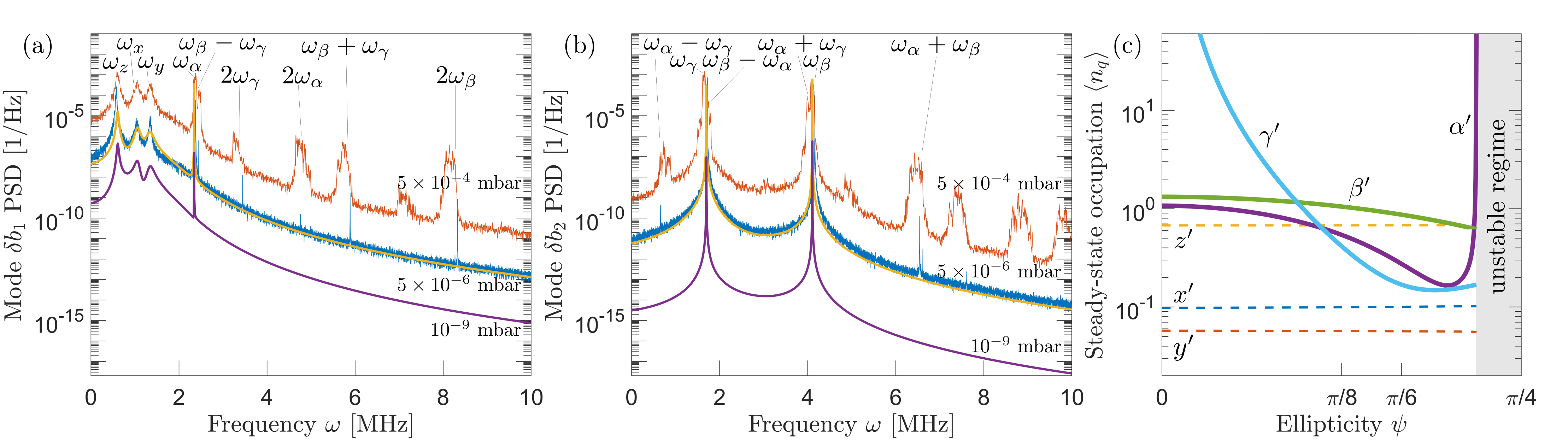}
    \caption{Analytic (thick solid line) and numerical (thin solid line) steady-state power spectral densities (PSDs) of the cavity modes $\delta b_1$ (a) and $\delta b_2$ (b). The PSDs are displayed for the three different gas pressures $p_{\rm g} = 5 \times 10^{-4}$\,mbar, $p_{\rm g} = 5 \times 10^{-6}$\,mbar, and $p_{\rm g} = 10^{-9}$\,mbar (from top to bottom). The additional peaks found numerically at high pressures are higher harmonics of the ground modes generated by the non-linearity of nanoparticle rotations. All other parameters are chosen as for (69,70,71)\,nm in Tab.~\ref{tab:occ}. (c) Steady-state occupations as a function of the tweezer ellipticity $\psi$ for the (69,70,71)\,nm particle in Tab.~\ref{tab:occ}.}
    \label{fig:2}
\end{figure*}

{\it Recoil heating---} So far, we discussed the conservative interaction between the mechanical degrees of freedom and the cavity modes. Recoil heating by incoherently scattered tweezer photons and collisional heating by residual gas atoms limits the cooling process. For recoil heating, the resulting  diffusion rates can be calculated either in a semiclassical picture \cite{seberson2019} or from the Lindbladians of the exact quantum master equation \cite{stickler2016part2}. For the center-of-mass modes $q \in \{x,y,z\}$, this yields the phonon heating rates \cite{prasubmission}
\begin{align} \label{eq:xi}
\xi_q^{\rm rec} = & \frac{\gamma_{\rm sc}\epsilon^2}{5}k^2 q_{\rm zp}^2\left[ (\chi_c^2 \cos^2\psi + \chi_b^2\sin^2\psi)\left (2+ u \,\delta_{zq} \right ) \right. \nonumber \\
 & \left. - \chi_c^2 \cos^2\psi \,\delta_{xq}- \chi_b^2\sin^2\psi \,\delta_{yq} \right],
\end{align}
with $u = 5(1-1/kz_{\rm R})^2$. A similar calculation for the librational degrees of freedom $q \in \{\alpha,\beta,\gamma\}$ yields the diffusion rates
\begin{align} \label{eq:xirot}
\xi_q^{\rm rec} = \gamma_{\rm sc}\epsilon^2 q_{\rm zp}^2\Delta\chi_q^2 \left[ 1 - \sin^2\psi \, \delta_{\beta q} - \cos^2\psi \, \delta_{\gamma q} \right],
\end{align}
where $\Delta\chi_\alpha = |\chi_b - \chi_c|$ and cyclic permutations.

The corresponding diffusion rates due to collisions with residual gas atoms are $\xi_q^{\rm gas} =  k_{\rm B} \gamma_q^{\rm gas} T_{\rm g}/\hbar \omega_q$, where $T_{\rm g}$ is the gas temperature and $\gamma_q^{\rm gas}$ denotes the gas damping constant \cite{martinetz2018}. {We neglect heating due to tweezer phase noise, which can be diminished by destructive interference between the stationary intra-cavity field $b_0$ and a phase-locked pump field.} The total heating rates are thus  given by $\xi_q = \xi_q^{\rm rec} + \xi_q^{\rm gas}$.

{\it Ground-state cooling---} The cavity-mediated coupling between the mechanical degrees of freedom leads to the appearance of hybrid mechanical modes $S_1' = \{x',y',z',\alpha'\}$ and $S_2' = \{\beta',\gamma'\}$, which can be determined by diagonalizing the first line of Eq.~\eqref{eq:ham2}. However, since the cavity population is much smaller than the tweezer amplitude, $|b_j| \ll \epsilon$, the couplings fulfill $|g_{qq'}|^2\ll \omega_q \omega_{q'}$ for most ellipticities, so that the hybridized modes $S_j'$ are well approximated by the original modes $S_j$ \cite{toros2019}.

After transforming to the hybrid modes, each mechanical degree of freedom $Q \in S_j'$ with trapping frequencies $\omega_Q$ linearly interacts with the respective light mode $b_j$ as quantified by the coupling constant $g_Q$. In the weak coupling approximation \cite{wilson2008}, the resulting cooling/heating rates due to enhanced anti-Stokes/Stokes scattering (Purcell effect) follow from a standard calculation as $\gamma_Q^{\mp} = 2|g_Q|^2\kappa/[\kappa^2 + (\Delta_j \pm \omega_Q)^2]$. These expressions are valid for $\gamma_Q^\mp \ll \kappa$ and $\gamma_Q^\mp \gamma_{Q'}^\mp \ll (\omega_Q - \omega_{Q'})^2$, where $Q,Q' \in S_j'$. From this, one obtains the stationary mechanical mode occupations $n_Q = (\gamma_Q^+ + \xi_Q)/(\gamma_Q^- - \gamma_Q^+)$. 

The resulting weak coupling steady-state occupations for three ellipsoidally shaped particles are listed in Tab.~\ref{tab:occ}. The corresponding cavity parameters and nanoparticle specifications are close to state-of-the art experiments \cite{delic2020}. Based on this we conclude that the 6D quantum ground state is realistically achievable by elliptic coherent-scattering cooling. Even though the rotation of an exactly spherical particle (first row) cannot be cooled, the orientation can be driven into the quantum ground state in the other cases considered. For increasingly anisotropic particles (second and third row) the librational and translational frequencies diverge, rendering simultaneous cooling inefficient. Nonetheless, appropriately choosing the detuning allows one to efficiently cool either the rotations or the center-of-mass motion to the  ground state. For slightly aspherical particles (fourth row), all six degrees of freedom can be simultaneously cooled into their quantum groundstate. {This regime can be reached in particular if all trapping frequencies lie within the cavity linewidth while being sufficiently separated to avoid dark modes. The impact of the particle anisotropy is discussed in Ref.~\cite{prasubmission}.} In Tab. \ref{tab:occ}, gas scattering is only relevant for the $\beta'$ and $\gamma'$ degrees of freedom in the fourth row. The corresponding damping constants are $\gamma_q^{\rm gas} \approx 5p_{\rm g}\ell_b^2 \sqrt{2\pi\mu}/6m\sqrt{k_B T_{\rm g}}$ \cite{martinetz2018} with $T_{\rm g}=300\,$K and the mass of helium $\mu$. 

The cooling timescales $\log(k_B T_0/\hbar\omega_Q n_Q)/(\gamma_Q^- - \gamma_Q^+)$ are on the order of a few hundred microseconds for the translation and in the two-digit millisecond regime for the librations, assuming a starting temperature of $T_0 = 40\,$K (librational trapping). Non-harmonicities in $H_0$  give rise to higher harmonics in the cavity output fields as long as the phonon number exceeds a few hundred. This is illustrated in Fig.~\ref{fig:2}, which shows the power spectral densities of the two cavity output modes for three different gas pressures \cite{prasubmission}. Figure~\ref{fig:2}(c) presents the steady-state occupations as a function of tweezer ellipticity $\psi$, demonstrating that ground-state cooling is impossible for linear ($\psi=\ 0$) and circular ($\psi=\ \pi/4$) polarization, and that the optimal cooling regime is close to $\psi=\ \pi/6$. Orientational trapping becomes unstable when approaching circular polarization since the time-averaged potential can no longer confine the rotation in the polarization plane.

{\it Discussion---} Coherent-scattering cooling with an elliptically polarized tweezer offers an attractive setup for efficiently cooling nanoparticles into the quantum regime in all their ro-translational degrees of freedom. This has great potential for sensing applications and for fundamental quantum experiments, even with spherical objects since exact sphericity can never be guaranteed.

Specifically, torque sensing can be best performed by monitoring the mode $b_2$, which is unaffected by the center-of-mass motion. The minimally detectable torque from measuring the two librational degrees of freedom $\beta,\gamma$ is $N_{2}^{\rm min}/\sqrt{B} \approx \sqrt{4\hbar\omega_{\gamma'}I_c \xi_{\gamma'}} \approx 3.9\times 10^{-30}\,\text{Nm}/\sqrt{\text{Hz}}$, where $B$ is the measurement bandwidth \cite{yin2013}. This would improve current experiments by orders of magnitude \cite{ahn2020} and enable the observation of rotational quantum friction \cite{zhao2012rotational} and the Casimir torque \cite{xu2017detecting}. Simultaneously monitoring both cavity output modes also opens the door to combined force and torque measurements using different degrees of freedom of a single levitated object.

\begin{figure}
    \centering
    \includegraphics[width = \columnwidth]{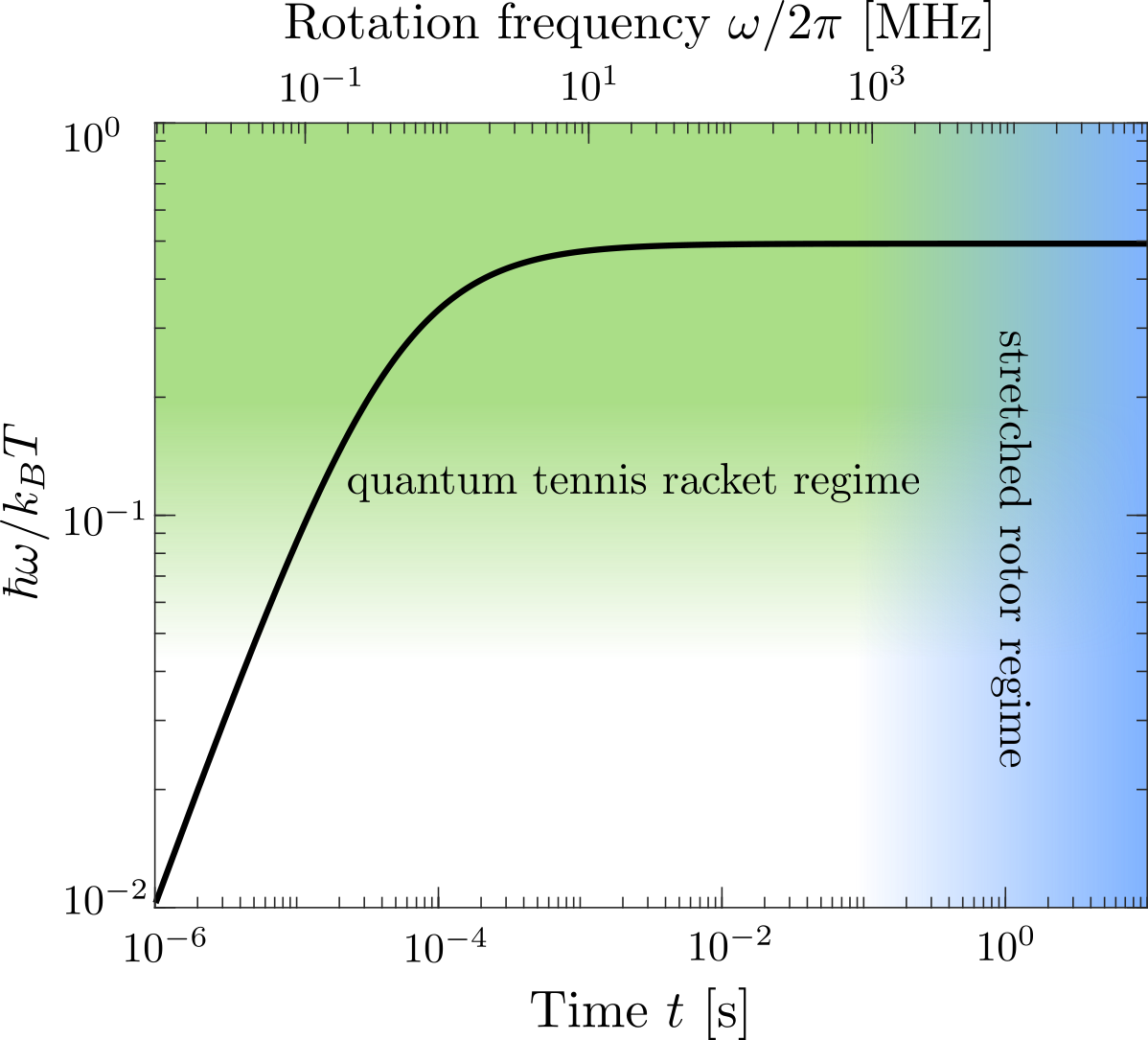}
    \caption{Rotationally cold nanoparticles can be propelled into the regimes where quantum tennis-racket flips \cite{ma2020} and the effects of centrifugal stretching \cite{hummer2020} are observable. The tennis racket regime is reached for $\hbar \omega/k_{\rm B} T \gtrsim 0.1$, where the rotational temperature $T$ characterizes the angular momentum width. Centrifugal stretching of the particle starts playing a role once the nanoparticle endpoint velocities become comparable to the speed of sound divided by its dielectric permittivity.}
    \label{fig:3}
\end{figure}

Even for particle shapes where 6D cooling is inefficient, elliptic coherent scattering can prepare their orientational degrees of freedom in the  quantum regime. This setup is therefore ideally suited for preparing rotational quantum superposition tests with nanoscale rotors \cite{stickler2018,ma2020}. For instance, cooling an asymmetric rotor into its quantum groundstate and then far-detuning and circularly polarizing the tweezer, so that the particle is angularly accelerated by the non-conservative torque \eqref{eq:torque}, generates a cold but rapidly rotating state as required to observe quantum persistent tennis racket flips \cite{ma2020} and acousto-rotational coupling \cite{hummer2020}. For the aspherical particles considered Tab.~\ref{tab:occ}, GHz rotation frequencies can be readily achieved by switching to circular polarization and detuning the tweezer from the cavity resonance. For instance, the (25,40,100)\,nm particle (Tab.~\ref{tab:occ}), initially prepared in the steady state and then rotationally accelerated by the non-conservative torque \eqref{eq:torque}, reaches the quantum tennis racket regime \cite{ma2020} in a matter of milli-seconds, see Fig.\,\ref{fig:3}.

Finally, the isolated librational $S_2'$ modes are ideally suited for trapped quantum experiments \cite{romero2011b,rudolph2020,rakhubovsky2020} because they are weakly affected by Rayleigh scattering of tweezer photons \eqref{eq:xirot}, leading to coherence times larger than that of the other mechanical modes by about one order of magnitude. For instance, the expected $\gamma'$ coherence time is on the order of $0.6 - 1.1$\,ms for the 6D ground state setup in Tab.~\ref{tab:occ}.

In summary, coherent-scattering cooling with an elliptically polarized tweezer enables simultaneous ro-translational ground-state cooling of nanoscale dielectrics. This setup may well serve as a building block for future quantum experiments  and sensors with levitated nanoparticles.

\begin{acknowledgments}
{\it Acknowledgments---} JS and HR contributed equally. We thank Stephan Troyer for helpful discussions and Issam Bouchdad for assistance in creating Fig.~\ref{fig:sketch54}. KH acknowledges funding by the Deutsche Forschungsgemeinschaft (DFG, German Research Foundation)--394398290, BAS acknowledges funding from the European Union’s Horizon 2020 research and innovation programme under the Marie Sk\l odowska-Curie grant agreement No. 841040  and by the Deutsche Forschungsgemeinschaft (DFG, German Research Foundation)--439339706.
\end{acknowledgments}

\end{document}